\documentclass[aps,prl,twocolumn,groupedaddress,superscriptaddress,showpacs]{revtex4}

\usepackage{amsmath}
\usepackage{amssymb}
\usepackage{graphicx}
\usepackage{dcolumn}
\usepackage{bm}
\usepackage[usenames]{color}
\usepackage[normalem]{ulem}
\usepackage[none]{hyphenat}

\begin{document}

\title{New topological surface state in layered topological insulators: unoccupied Dirac cone}

\author{S.V. Eremeev}
 \affiliation{Institute of Strength Physics and Materials Science,
634021, Tomsk, Russia}
 \affiliation{Tomsk State University, 634050 Tomsk, Russia}

\author{I.V. Silkin}
 \affiliation{Tomsk State University, 634050 Tomsk, Russia}

\author{T.V. Menshchikova}
 \affiliation{Tomsk State University, 634050 Tomsk, Russia}

\author{A.P. Protogenov}
 \affiliation{Institute of Applied Physics, Nizhny Novgorod, 603950, Russia}

\author{E.V. Chulkov}
\affiliation{Donostia International Physics Center (DIPC),
             20018 San Sebasti\'an/Donostia, Basque Country,
             Spain\\}
\affiliation{Departamento de F\'{\i}sica de Materiales UPV/EHU,
Centro de F\'{\i}sica de Materiales CFM - MPC and Centro Mixto
CSIC-UPV/EHU, 20080 San Sebasti\'an/Donostia, Basque Country, Spain}

\begin{abstract}
The unoccupied states in topological insulators Bi$_2$Se$_3$,
PbSb$_2$Te$_4$, and Pb$_2$Bi$_2$Te$_2$S$_3$ are studied by the
density functional theory methods. It is shown that a surface state
with linear dispersion emerges in the inverted conduction band
energy gap at the center of the surface Brillouin zone on the (0001)
surface of these insulators. The alternative expression of
$\mathbb{Z}_2$ invariant allowed us to show that a necessary
condition for the existence of the second $\bar\Gamma$ Dirac cone is
the presence of local gaps at the time reversal invariant momentum
points of the bulk spectrum and change of parity in one of these
points.
\end{abstract}

\pacs{73.20.-r, 71.70.Ej, 72.25.Dc}

\maketitle

\begin{figure*}
\includegraphics[width=\textwidth]{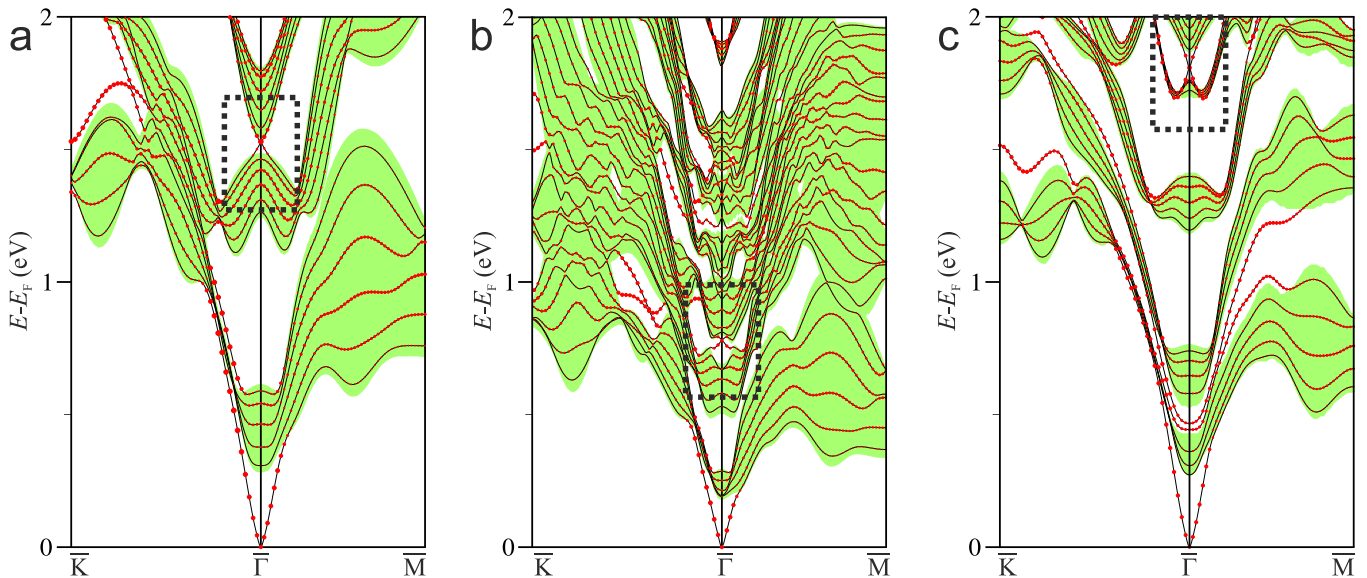}
 \caption{Fig.~1 Surface electronic structure of Bi$_2$Se$_3$ (a), PbSb$_2$Te$_4$ (b),
 and Pb$_2$Bi$_2$Te$_2$S$_3$ (c) above $E_{\rm F}$  as calculated by VASP.
 Size of red circles correspond to the weight of the states in the
 outermost block.  The projected bulk bands is shown in green.
 Dashed frames mark region of the unoccupied Dirac-like surface state.}
 \label{fig1}
\end{figure*}

Three-dimensional topological insulators (TIs) are characterized by
presence of the topological surface state (SS) emergence in the
principal band gap \cite{Zhang_NatPhys2009}. This spin-polarized
linearly dispersing surface state forming Dirac cone arises from a
symmetry inversion of the bulk bands at band gap edges owing to the
strong spin-orbit interaction (SOI). Time reversal symmetry protects
topological SS from backscattering in presence of weak perturbation
\cite{FKM_2007,Qi_2008, Roushan_Nat2009}. This causes new
possibility for practical applications particularly in realization
of dissipationless spin transport which can be used in new
spintronic devises.

A number of materials that hold non-trivial spin-polarized Dirac
state has been intensively studied. Among them the Bi and Sb
chalcogenides are most investigated TIs at present. Their band
structure is widely studied
\cite{Zhang_NatPhys2009,ChenYL2009Science,Hsieh_Nat2009,ZhangT2009PRL,Kuroda_PRL10a,Xu10,EremeevJETPL2010}
and spin texture of the topological SS was observed both indirectly
by use of circular dichroism \cite{Wang2011} and directly by
spin-resolved experiments \cite{Hsieh_Nat2009,Eremeev_NatCom,
PRL12_ternary}. However, besides the Dirac cone various types of SS
states take place in TIs. Some of them can result in new physical
phenomena. For example, angle-resolved photoemission spectroscopy
(ARPES) measurements as well as time- and angle-resolved
photoemission spectroscopy experiments \cite{Sobota} led to
discovery of parabolic spin-split surface states in the principal
energy gap just below the conduction band and M-shaped states in the
local valence band gap
\cite{Bianchi,Wray_NatPhys11,Valla,Wray_arXiv,Zhu_PRL,JetpLett_M11,
Benia_PRL,JETP_lett_M,NewJPhys}.

Here we study the conduction band (CB) energy gap surface states in
the layered topological insulators Bi$_2$Se$_3$, PbSb$_2$Te$_4$, and
Pb$_2$Bi$_2$Te$_2$S$_3$ by using density function theory (DFT)
methods. The choice of these TIs is motivated by the fact they have
different crystal structure, composed of quintuple (Bi$_2$Se$_3$),
septuple (PbSb$_2$Te$_4$), and nonuple (Pb$_2$Bi$_2$Te$_2$S$_3$)
layer blocks. Bi$_2$Se$_3$ is the extensively studied TI  while
PbSb$_2$Te$_4$ and Pb$_2$Bi$_2$Te$_2$S$_3$ were recently predicted
as TIs \cite{Eremeev_NatCom,Silkin}. We show that except the Dirac
cone in the principal gap and trivial unoccupied surface states a
massless spin polarized surface state arises in the CB local gap at
the $\bar\Gamma$. This state induced by the SOI inverted local gap
has spin helicity similar to that of the Dirac state in the
principal energy gap.

For electronic band calculations we employ two different computer
codes that are based on DFT. The first one is the Vienna Ab Initio
Simulation Package (VASP) \cite{VASP1,VASP2}. We used the
generalized gradient approximation (GGA) \cite{PBE} to the exchange
correlation potential and the projector augmented wave (PAW)
\cite{PAW1,PAW2} basis sets to solve the resulting Kohn-Sham system.
The second approach used for electronic structure calculations is
the full-potential linearized augmented plane-wave (FLAPW) method as
implemented in the FLEUR code \cite{FLEUR} with PBE for the
exchange-correlation potential. The FLAPW basis has been extended by
conventional local orbitals to treat quite shallow semi-core
$d$-states. Both methods contained scalar relativistic corrections
and spin-orbit coupling was taken into account by the second
variation method \cite{KH}. To simulate the (0001) surfaces of the
TIs we use a slabs composed of 6 quintuple layers (QLs) and 6
septuple layers (SLs) for Bi$_2$Se$_3$ and PbSb$_2$Te$_4$,
respectively, and 5 nonuple layers (NLs) for
Pb$_2$Bi$_2$Te$_2$S$_3$.

The calculated unoccupied electronic structure spectra in materials
under interest is shown in Fig.~\ref{fig1}. Shaded regions depict
projected bulk bands onto the (0001) plane, while red dots indicate
a weight of the states in the outermost structural block (QL, SL,
NL). As easy to see in addition to the Dirac cone in the principal
gap there exist different trivial surface states at the energies
shown in Fig.~\ref{fig1} and at higher energies in the local CB gaps
in all surfaces of interest.

\begin{figure}
\includegraphics[width=\columnwidth]{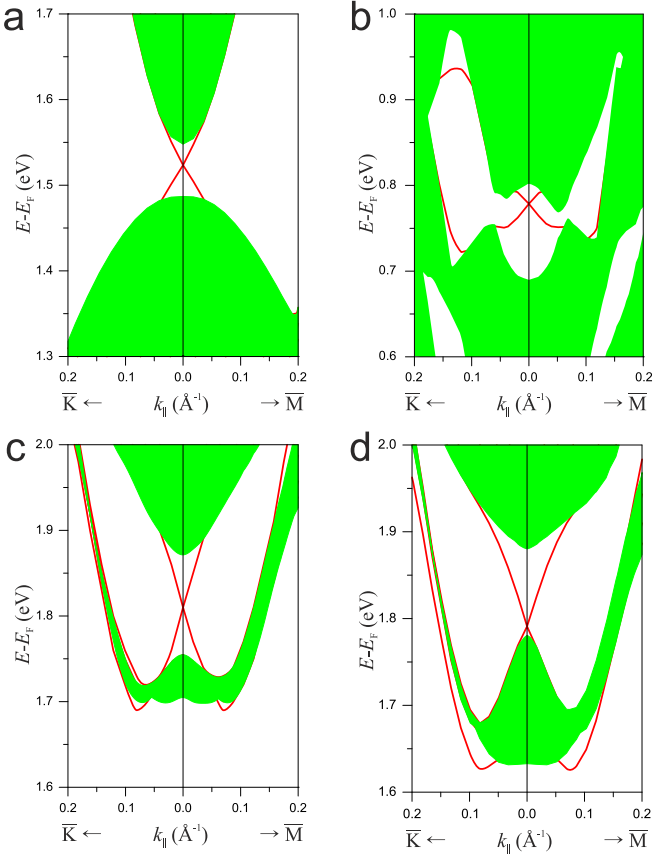}
 \caption{Fig.~2 A magnified view of dashed frames marked in the Fig.~\ref{fig1}
with CB cone in the center: (a) Bi$_2$Se$_3$; (b) PbSb$_2$Te$_4$;
Pb$_2$Bi$_2$Te$_2$S$_3$ as calculated by VASP (c) and FLEUR (d)
codes.}
 \label{fig2}
\end{figure}

At the same time in the conduction band energy gap there appears a
Dirac-like surface state at the $\bar\Gamma$ point (framed in
Figs.~\ref{fig1}(a-c) by dashed line) for all compounds we study. In
the case of Bi$_2$Se$_3$ this state arises at $\sim 1.5$ eV in gap
of 70 meV formed by projection of the second and third bulk
conduction bands. For PbSb$_2$Te$_4$ and Pb$_2$Bi$_2$Te$_2$S$_3$ the
Dirac-like surface state takes place at $\sim 0.8$ eV and $\sim 1.8$
eV respectively. These states have the linear dispersion in the
local gap and entering to the bulk states projection they transform
into resonances and spread along the edges of the projection. A
close-up of these states is shown in Fig.~\ref{fig2}. Note that the
Dirac-like surface state in the conduction band is well reproduced
by both VASP and FLEUR calculations with some differences in the
projected bulk bands (Figs.~\ref{fig2} (c) and (d)).

\begin{figure}
\begin{center}
\includegraphics[width=0.5\columnwidth]{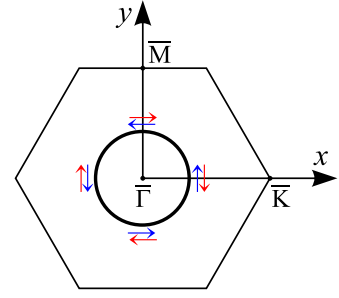}
\end{center}
 \caption{Fig.~3 Schematic view of the spin structure of the CB cone.
 Red and blue arrows indicate spin direction for upper and lower part of the
 cone, respectively. Scales for constant energy contour and 2D Brillouin zone are different
 (typical maximal radius of the contour in materials under study is $\sim 0.03$\AA$^{-1}$;
 $\bar\Gamma - \bar{\rm K}$ direction is of $\sim 1$\AA$^{-1}$).}
 \label{fig3}
\end{figure}

In all considered TIs the unoccupied Dirac-like SS is isotropic with
respect to $k_{||}$, i.\,e. the constant energy contours below and
above the degeneracy point have an ideal circular shape as
schematically shown in Fig.~\ref{fig3}. The calculated orientation
of the electron spin within the surface state demonstrates in-plane
spin polarization (out-of-plane component is negligibly small) with
positive (clockwise) spin helicity in the upper part and negative
helicity in the lower part of the Dirac cone, i.\,e. the same
helicity as in the case of the topological SS located in the
principal energy gap. Besides, owing to the SOI entanglement of the
spin and orbital momenta the spin polarization in the CB cone is
reduced and does not exceeds 60\% in Bi$_2$Se$_3$ and
Pb$_2$Bi$_2$Te$_2$S$_3$ and is slightly smaller in PbSb$_2$Te$_4$
($\sim$ 50\%) that is comparable with the spin polarization in the
conventional Dirac cone \cite{Yazyev, PRL12_ternary}.

\begin{figure}
\includegraphics[width=\columnwidth]{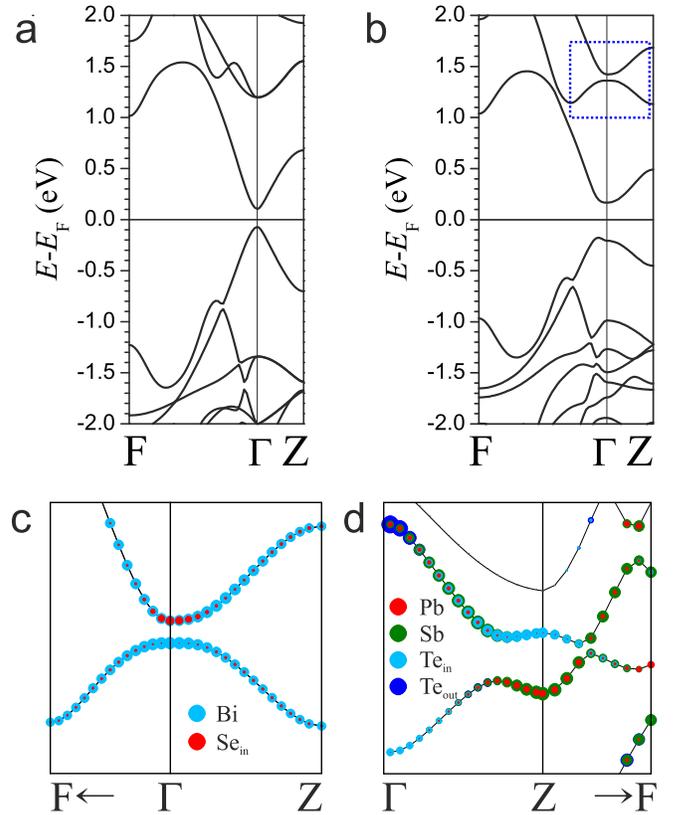}
 \caption{Fig.~4 Bulk band structure of Bi$_2$Se$_3$
 without (a) and with (b) SOI included; (c) magnified view of blue dashed frame marked in the
 Fig.~\ref{fig4}(b) with composition of the second and third
 conduction bands in the vicinity of $\Gamma$; (d) the same for PbSb$_2$Te$_4$ in the vicinity of Z.}
 \label{fig4}
\end{figure}

In order to reveal the origin of formation of the second Dirac state
in the conduction band of TIs consider first the bulk band structure
of Bi$_2$Se$_3$. Note that a direct consequence of the inversion of
the Bi and Se states in the principal gap at the $\Gamma$ point in
Bi$_2$Se$_3$ is formation of the Dirac cone in the principal energy
gap. Let us consider the bulk conduction band of Bi$_2$Se$_3$ in the
energy range of interest. Without spin-orbit coupling included the
second and third conduction bands are degenerate along the $\Gamma -
{\rm Z}$ direction (Fig.~\ref{fig4}(a)). Spin-orbit interaction
lifts this degeneracy and opens a gap between these bands at the
$\Gamma$ point (Fig.~\ref{fig4}(b)). Thus, the gap supporting the
new massless state in the conduction band results from the SOI. Both
the upper and lower bands are mostly composed of the bismuth states,
however, the Se $p_z$ states of the central atomic layer of QL
contribute to these bands too. As one can see in Fig.~\ref{fig4}(c),
the weight of Se states in the lower band goes to zero approaching
the $\Gamma$ point. At the same time Se states in the upper band
have maximal weight in the vicinity of $\Gamma$. Thus, this
SOI-induced local gap at $\sim 1.5$ eV between the second and third
conduction bands has inverted Se states. Like to the principal
$\Gamma$ gap where the states of the outer Se atoms of QL are
inverted the states of the inner Se atom have a similar SOI-induced
inversion in the CB gap. This fact points out that the CB Dirac cone
is the unoccupied topological SS.

In the bulk band spectra of PbSb$_2$Te$_4$ the similar SOI-induced
inversion can be detected for states of the inner Te atoms of SL in
the vicinity of the Z point (Fig.~\ref{fig4}(d)). However, in this
ternary compound the picture is more complicated and it becomes even
less obvious for quaternary TI Pb$_2$Bi$_2$Te$_2$S$_3$.

From this point of view it would be desirable to have such a simple
technique to qualify unoccupied topological states as calculation of
${\mathbb Z}_2$ invariant. The difficulty is that the unoccupied
cone is appeared in the local gap of the bulk conduction band where
along a path between time-reversal invariant momenta (TRIMs) in the
Brillouin zone the gap in the spectrum of the bulk states closes and
opens.

Analysis of the relation $(-1)^{\nu_{0}}=(-1)^{2P_{3}}$ \cite{WQZ}
of the index $\nu_{0}$ \cite{FKM_2007,FK2007} in the theory of
topological insulators with the winding number $2P_{3}$ or with the
non-Abelian Chern-Simons term in the $k$-space \cite{QHZ} using the
continuous approach in the theory of topological invariants does not
solve the problem. Thus we address the approach of Refs.
\cite{FK,FH,U} enabling us to determine the mechanism of reduction
of the contribution made by bulk states to the topological
invariant.

The origin of reduction of the contribution of the bulk states can
be found by comparing the continuous and lattice versions of the
$\mathbb{Z}_2$ invariant. It is shown in Ref. \cite{FK} that the
$\mathbb{Z}_2$ invariant in the continuous case is expressed as
\begin{alignat}{1}
D&=\frac{1}{2\pi i}\left[ \oint_{\partial{\cal B}^-}A-\int_{{\cal
B}^-} F \right]\,\,\, {\text{\rm mod 2} }, \label{FuKane}
\end{alignat}
where ${\cal B}^-=[-\pi,\pi]\otimes[-\pi,0]$ is half of the
Brillouin zone, $A=Tr \, \psi^{\dagger}d\psi$ and $F=dA$ are the
Berry gauge potential and the associated field strength respectively
and $\psi (k)$ is the 2M(k)-dimensional ground state multiplet. The
lattice analog of Eq. (\ref{FuKane}) is \cite{FH}
\begin{alignat}{1}
D_{\rm L} &\equiv\frac{1}{2\pi i}\left[ \sum_{k \in\partial{\cal
B}^-}A(k) -\sum_{k \in{\cal B}^-}F(k) \right]
\nonumber\\
&=-\sum_{k \in{\cal B}^-}n(k)\,\,\, {\text{\rm mod 2}} \, ,
\label{FuHat}
\end{alignat}
since $\sum_{k \in {\cal B}^-}F(k)= \sum_{k \in\partial{\cal
B}^-}A(k) +2\pi i\sum_{k \in{\cal B}^-}n(k)$. $n(k)$ in this
equation are integers and $n(k) \,{\rm mod \, 2} \in \mathbb{Z}_2$
due to the residual $U(1)$ invariance \cite{FH}. From Eq.
(\ref{FuHat}) we can conclude that the reason for cancelation of the
bulk continuum states is the compactness of the lattice gauge
theory. In all TIs under study at energy of CB Dirac state there are
gaps in $\Gamma$, F$_i$, L$_i$, and Z TRIMs in the case of
rhombohedral Bi$_2$Se$_3$ and PbSb$_2$Te$_4$ or in $\Gamma$, M$_i$,
L$_i$, and A TRIMs of hexagonal Pb$_2$Bi$_2$Te$_2$S$_3$ which allow
us to implement the parity analysis \cite{FK2007} to calculate the
topological number $\nu_0$ for the CB cone in these systems by
fixing of an isoenergy level in the local gap, supporting the
unoccupied Dirac cone.

The results for PbSb$_2$Te$_4$ show that non-trivial value $\nu_0=1$
of the topological invariant for the unoccupied cone is provided by
the change of the parity at the Z point (where we have revealed the
inversion of Te$_{in}$ states, see Fig.~\ref{fig4}(d)). In the case
of Pb$_2$Bi$_2$Te$_2$S$_3$ the change of parity in the A point in
the gap between 4-th and 5-th bulk conduction bands determines
topologically non-trivial $\nu_0$ and the appearance of the
unoccupied Dirac cone in the surface electronic spectrum. Thus
inversion of the local gap edges in the conduction band along with
the presence of non-inverted gaps in other TRIMs of the bulk
Brillouin zone at the same energy is responsible for the emergence
of single Dirac-like helical spin surface state in the topological
insulators.

In summary, we have examined the unoccupied electronic spectra of
the layered topological insulators with different composition and
crystal structure. We revealed that except for trivial conduction
band surface states a massless topologically-protected helical spin
state exists in the narrow local conduction band gap at the
$\bar\Gamma$ point in all materials under study. This state arises
owing to inversion of the certain bulk conduction bands at the
time-reversal invariance momentum points $\Gamma$, Z, and A of
Bi$_2$Se$_3$, PbSb$_2$Te$_4$, and Pb$_2$Bi$_2$Te$_2$S$_3$,
respectively. The observed state has the same spin helicity as in
the conventional Dirac cone in the principal energy gap. The
revealed CB topological states provide a pathway to the measurements
on excitations between two Dirac spin polarized states at the
surface of layered topological insulators.

\end{document}